# A study of event mixing for two-pion Bose-Einstein correlations in the $\gamma p \rightarrow \pi^0\pi^0 p$ reaction

Q. He[1,2,*], J. Ai[1], T. Ishikawa[2], T. Li[1], L. Ma[1], J. Ma[1], M. Miyabe[2], N. Muramatsu[2], H. Shimizu[2], Y. Tsuchikawa[2], Y. Xiang[1], H. Yamazaki[2], Y. Zhang[1]

[1] *Institute of Fluid Physics, China Academy of Engineering (CAEP), P. O. Box 919-101, Mianyang 621900, China*
[2] *Research Center for Electron Photon Science, Tohoku University, Sendai 982-0826, Japan*

**E-mail:* hetsinghua@caep.cn.

**Abstract**

We develop an event mixing technique with a missing mass consistency (MMC) cut and a pion energy (PE) cut for the measurement of two-pion Bose-Einstein correlations (BEC) in reactions with only two identical pions produced in the final state. A multi mixing mode that allows one event to be mixed with multiple other events is proposed for the purpose of improving the statistics of mixed samples, and compared with the single mixing mode which requires one original event can be mixed only once. Numerical tests using the $\gamma p \rightarrow \pi^0\pi^0 p$ events are used to testify its ability to measure two-pion correlations. To improve the mixing speed, the optimal MMC cut window is also explored via simulations.

## 1. Introduction

The Bose-Einstein correlations (BEC) can be used as a tool to investigate the space-time properties of subatomic reactions emitting identical bosons [1]-[3]. This technique involves the measurement of a two-particle correlation function defined as:

$$C_{BEC}(p_1,p_2) = \frac{P_{BEC}(p_1,p_2)}{P_0(p_1,p_2)} = 1 + \left|f(q)\right|^2 , \quad (1)$$

where $P_{BEC}(p_1,p_2)$ stands for the probability density of two identical bosons with momenta $p_1$ and $p_2$ radiated from a spatially localized source, $P_0(p_1,p_2)$ represents the emission probability of the so-called "reference sample" in the absence of BEC effects. The reference sample is generally produced through event mixing technique [20], which eliminates the BEC effects through selecting two bosons' momenta from different events. The quantity $f(q)$ is the Fourier transform of the emitter source distribution, where $q = p_1 - p_2$. If a sphere of source with a Gaussian density distribution is assumed, Eq. (1) is written as [4]

$$C_{BEC}(p_1,p_2) = C_{BEC}(Q) = N(1 + \lambda_2 e^{-r_0^2 Q^2}) \quad (2)$$

where $Q$ stands for the relative momentum of two bosons defined by $Q^2 = -q^2 = M^2 - 4\mu^2$ ($M$ is the invariant mass of the two identical bosons of mass $\mu$), and $r_0$ denotes the Gaussian parameter corresponding to the size of the emitter source. $N$ is the normalization factor. The parameter $\lambda_2$, reflecting the chaoticity of the source, is a measure of the BEC strength ranging from 0 to 1, where 0 and 1 correspond to completely coherent case and totally chaotic emission, respectively. Experimental factors such as particle misidentification and detecting resolutions have an influence on the measurement of bosons' correlations and consequently are reflected in $\lambda_2$.

BEC has a widespread application in the investigation of the space-time properties of a reaction volume for high-energy elementary-particle collisions [5]-[15] and relativistic heavy-ion collisions [16]-[19] with a large multiplicity. However, BEC observation for exclusive reactions with only two identical bosons at low energies is still one of the big challenges, which may offer complementary information compared to inclusive reactions. It can reveal the spatial and temporal characteristics of baryon resonances excited by hadronic or electromagnetic probes in the non-perturbative QCD energy region. Analyzing the correlations of identical bosons emitted from two excited baryons will yield information on the duration and the size of one of the excited baryons.

The event mixing method, which works at a certain level at high energies with a large multiplicity, is strongly disturbed by non-BEC factors of exclusive reactions with a low multiplicity such as global conservation laws and decays of resonances [21][22]. Significant kinematical correlations of final state particles induced by conservation laws complicate the mixing [23][24]. Poor understanding of these effects impedes the BEC investigation for exclusive reactions. In the work of [25], they investigated the effects of kinematical correlations on the event mixing for the $\gamma p \to \pi^0\pi^0 p$ reaction at incident photon energies $E_\gamma$ around 1 GeV (a non-perturbative QCD region). An event mixing method containing two kinematical cuts was proposed. The first cut, named missing mass consistency (MMC) cut was used to conserve the energy momentum, and to make physically meaningful mixed events. The second cut, named pion energy (PE) cut was used to tune the slope of the correlation function in the hope of effective extraction of BEC parameters. Several numerical tests validated its efficiency and correctness.

Because the PE cut requires excluding a certain group of original events [25], the statistical error becomes worse. Although the single mixing strategy, which requires one original event can be mixed only once, guarantees equal contributions from every event, it is not a good choice as long as the statistical error is concerned. In this work, a multi mixing mode is proposed to solve this problem. It allows one event to be mixed with multi events and hence improves the statistics of the mixed events. Results from the single and multi mixing modes are compared. The ability of the multi mixing mode to observe BEC effects is validated and compared with the single mixing mode. Good agreement is found between them.

To improve the statistical error, a possible way is using events from different system energy bins together. This idea is however conflict with the prerequisite of the MMC cut, which requires two events from the same system energy bin can be mixed. To test the tolerance of this prerequisite to the energy span, we also investigate different widths effects on the correlation function.

The cut window width of the MMC cut has a significant impact on the shape of the correlation function and consequently influences the accuracy of BEC observation. It also has an influence on the event mixing speed. An adequate window that not only constructs a valid correlation function but also takes a shorter mixing time is useful. In this work, we also investigate the MMC cut window effects on the event mixing.

The paper is organized as follows. In Sect. 2, we describe the simulation method of the multi mixing. In Sect. 3, extensive numerical tests which simulate the application of the multi mixing mode to the $\gamma p \to \pi^0\pi^0 p$ reaction are presented. In the simulation, we investigate the MMC cut window effects, beam energy width effects and the ability to observe BEC parameters. In Sect. 4, a summary is given finally.

## 2. Simulation method

The reference sample is produced based upon the original event sample through event mixing via taking two bosons from two randomly selected events under prescribed cut conditions. Correspondingly, a Monte Carlo program named Event Mixing for Bose-Einstein correlations (EMBEC) is developed. In order to improve the random event mixing speed, this program optimize the data access via using ROOT data formatting strategies because it is able to read inserted data without reading the entire event every time. The TTree [26] class in ROOT is adopted in EMBEC to store data of events due to its advantages of reducing disk space and of enhancing access speed. Those advantages are achieved through writing data a bunch at a time and reading individual variables independently.

The EMBEC program mainly contains four steps. In the first step, the whole sample should be trimmed through eliminating events that do not meet prescribed requirements of event mixing. In the second step, the trimmed sample is divided into several sub-samples if necessary. Event mixing is performed independently in every subsample for a group of classified events such as those grouped in terms of the mass of the remaining baryon in the final state. This treatment enables to largely reduce the time for mixing because it avoids scanning data over the whole sample. In the third step, event mixing runs under certain cut conditions. Mixed events failing to satisfy these conditions are rejected. In the final step, mixed events from all sub-samples are combined together to construct a mixed sample.

In the mixing process, a particle from the $i$-th event ($1 \leq i \leq n$) are paired with another particle selected via scanning the remaining list of events available to be a mixed event, which is tested whether it satisfies all cut conditions. The scanning begins with the event of index $j$ determined by:

$$j = (i + R) \bmod n, \text{ and } i \neq j , \tag{3}$$

where $R$ is a random integer uniformly distributed in the region [0, $n$-1]. The scanning stops when all the remained $n-1$ events are scanned or the number of scanned events exceeds the prescribed maximum value $n_{scan}^{max}$. This random scanning can effectively eliminate possible correlations between the mixed events.

There are two strategies for the pairing method: (1) single mixing; (2) multi mixing. In the single mixing mode, each event is used only once. Although this mode guarantees equal contributions from every event, it limits the statistics of the mixed events. In contrast, the multi mode allows one event to be mixed with multi events and hence improves the statistics of the mixed events. In the single mode, two events can produce a maximum of two mixed events with certain prescribed constraints. To keep an equal event contribution, the weight of a mixed event is set to be 1 (1/2) when just one (two) mixed events are produced. Events failing to be paired under certain cuts will not be included in the original sample when calculating the correlation function. In the multiple mode, one event can produce a maximum of $n_{mix}^{max}$ mixed events. The value of $n_{mix}^{max}$ should be optimized, by taking into consideration the time consumption and the statistical factor of mixed events. If one event totally makes $n_{mix}$ mixed events, the weight of each mixed event is set to be $1/n_{mix}$.

## 3. Numerical test

We apply the proposed multi-mixing method to the $\gamma p \to \pi^0\pi^0 p$ reaction to investigate MMC cut window effects. In the simulation, a ROOT utility named "TGenPhaseSpace" [27] is used to generate random $\gamma p \to \pi^0\pi^0 p$ events following the pure three-body phase space distribution. To generate an event, a total energy ($E_{tot}$), a multiplicity ($N$) and a list of masses ($m_i$) of emitted particles are required as input parameters. For each event, it returns a weight proportional to the naturally appearing probability. This weight is based upon the phase-space integral $R_N$

$$R_N = \int^{4N} \delta^4(P - \textstyle\sum_{j=1}^{N} p_j) \prod_{i=1}^{N} \delta(p_i^2 - m_i^2) d^4 p_i, \tag{4}$$

where $P$ and $p_i$ are the four momentum of the whole system and that of individual emitted particles, respectively.

Two constraints are employed in the event mixing: the MMC cut and the PE cut [25]. The MMC cut rejects such a mixed event as does not satisfy the relation $\left|m_X^{mix} - m_x^{ori}\right| < M_{cut}$ where $\mathrm{m_X^{mix}}$ and $\mathrm{m_x^{ori}}$ are the missing particle mass in the mixed event $\pi^0\pi^0\mathrm{X}$ and that in the original event, respectively. $M_{cut}$ is the cut window. It successfully forces the mixed events in the allowed phase space region. The prerequisite of this cut is that the $m_x^{ori}$ values of two mixing original events should be close enough. The four-momentum of the missing particle $P_X$ is calculated through four-momentum conservation: $P_X = P - P_1 - P_2$ , where $P$ is the four-momentum of the total system, and $P_1$and $P_2$ are the four-momenta of two mixed pions. The PE cut is able to produce a sample of mixed events identical to the original events as long as the event density distribution in terms of $Q$ is concerned. The PE cut requires the energies of particles used for mixing must not exceed a given value $E_{max}$. For the $\gamma\mathrm{p} \rightarrow \pi^0\pi^0\mathrm{X}$ reaction $E_{max}$ is found to depend on the incident photon energy $E_\gamma$ and the invariant mass $m_X$ of the missing particle $X$ [25]:

$$E_{max}/E_\gamma \ = 1.1 - 0.64 m_X, \tag{5}$$

where $m_X$ is given in the unit of GeV.

With these two cuts, event mixing is performed for the $\gamma\mathrm{p} \rightarrow \pi^0\pi^0\mathrm{p}$ events generated in pure phase space at an incident photon energy of 1.0 GeV. Fig. 1 shows the correlation functions obtained in the single mixing mode and the multi-mixing mode which produces up to 10 mixed events for one original event. It is found that the single mixing and multiple mixing modes give consistent correlation functions within the error bars as long as the slope of the correlation function is concerned. Fig. 2 shows the distributions of the number of scanning events in event mixing under the MMC cut and PE cut when the maximum mixing number is set to be $n_{scan}^{max}$=60, 100 or 140, respectively. In the simulation, the MMC cut window is set to be $M_{cut}$ =5 MeV. It is found totally 12%, 4%, and 2% of original events fail to be paired at $n_{scan}^{max}$=60, 100, and 140, respectively. If CPU time is not a crucial factor, a larger $n_{scan}^{max}$ value is recommended because it provides high statistics for mixed events.

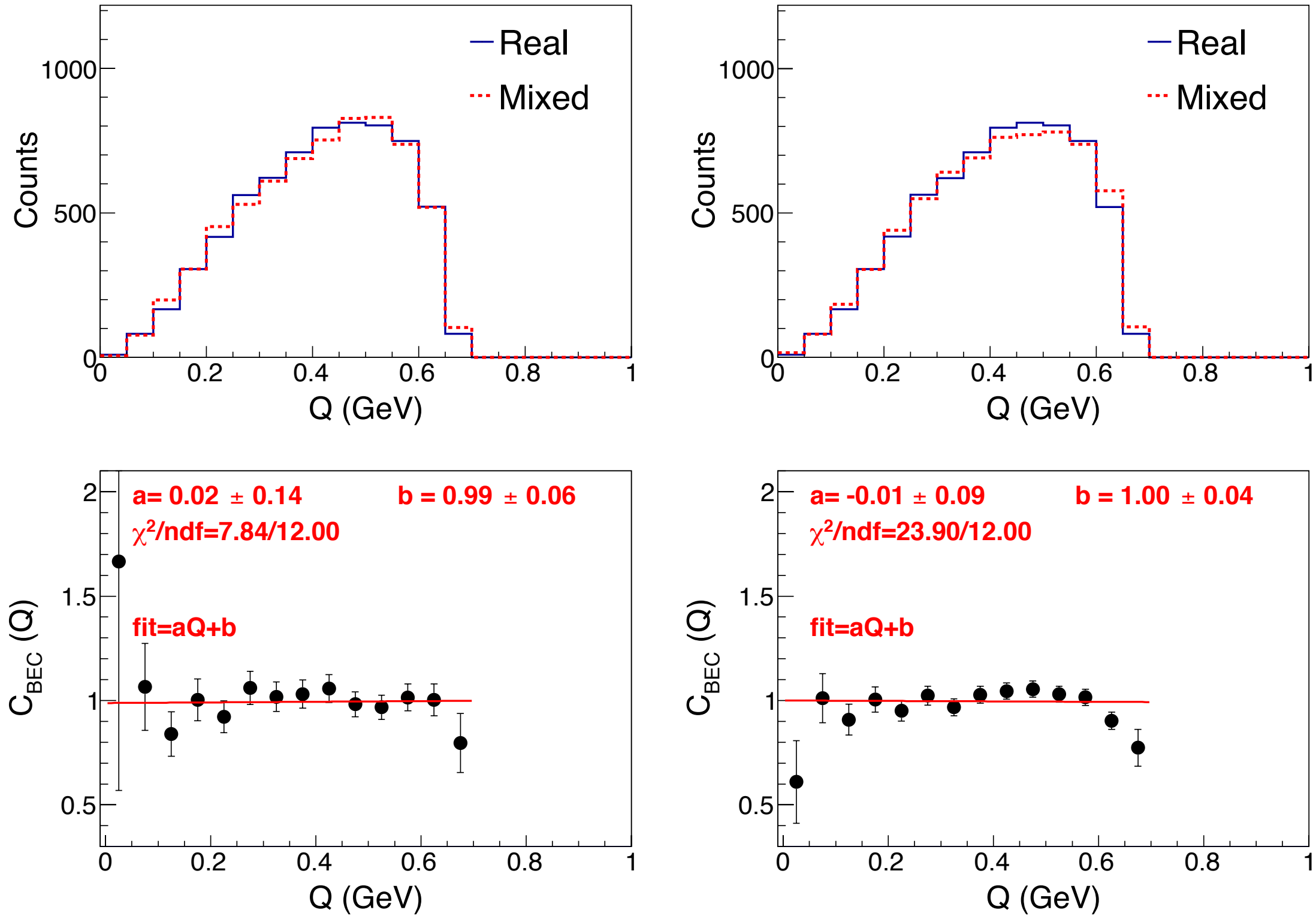

Fig. 1. $\mathbf{Q}$ distributions (upper) and the correlation functions (lower) obtained by event mixing in single mixing mode (left) and multi-mixing mode (right) for $\gamma p \to \pi^0\pi^0 p$ events at an incident photon energy of $E_\gamma$=1.0 GeV. A linear function *f(Q)=aQ+b* is fitted to the correlation functions.

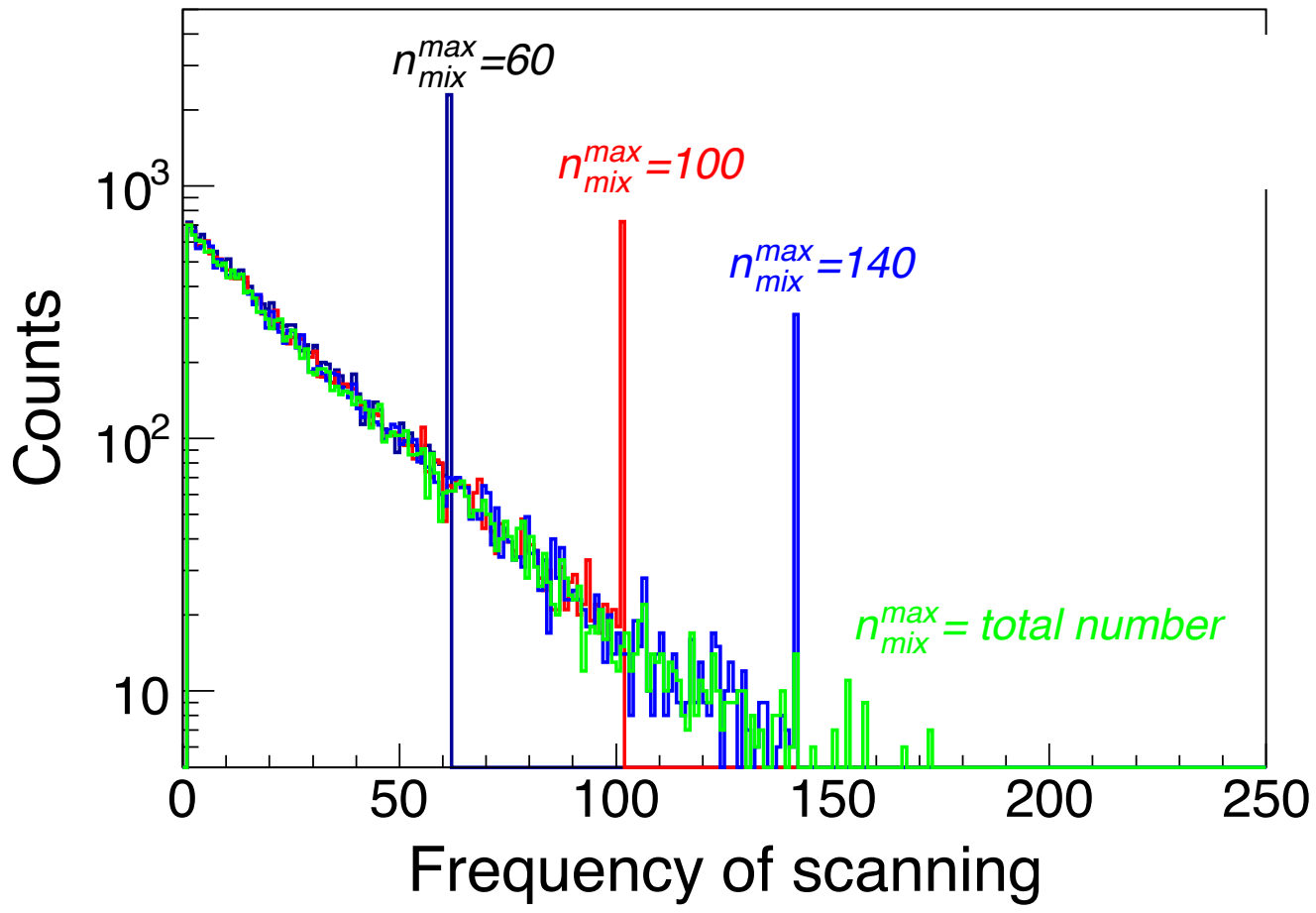

Fig. 2. Number of scanning events in pairing events when the MMC cut and the PE cut are employed in event mixing.

The cut window width of the MMC cut brings a significant impact on the shape of the correlation function, and also on the scanning frequency of finding partners in event mixing. To investigate this effect, several different cut window widths are tested. Fig. 3 shows three correlation functions obtained at three typical MMC cut window widths of 0.005, 0.1, and 0.2 GeV, respectively. The slope of the correlation function increases as the cut window width gets wider.

When the window width goes down to 30 MeV, the slope reaches 0 and stays around 0 at lower cut window values, as shown in Fig. 4. It is also found that the maximum frequency of event scanning increases with the cut window becoming narrower (see Fig. 5).

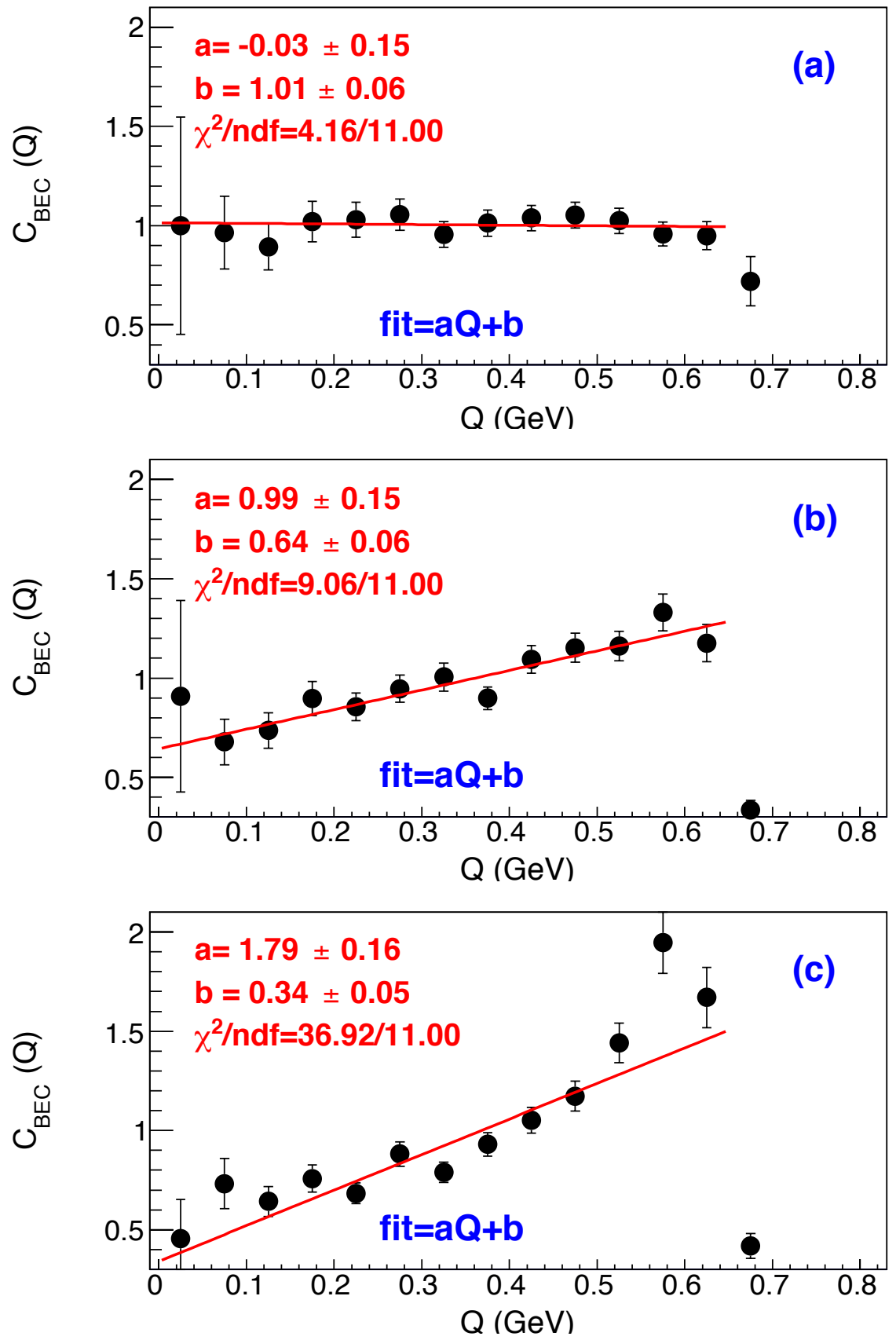

Fig. 3. Correlation functions obtained with different MMC cut window widths of 0.005 GeV (a), 0.1 GeV (b), and 0.2 GeV (c), respectively. A linear function *f(Q)=aQ+b* is used in fitting.

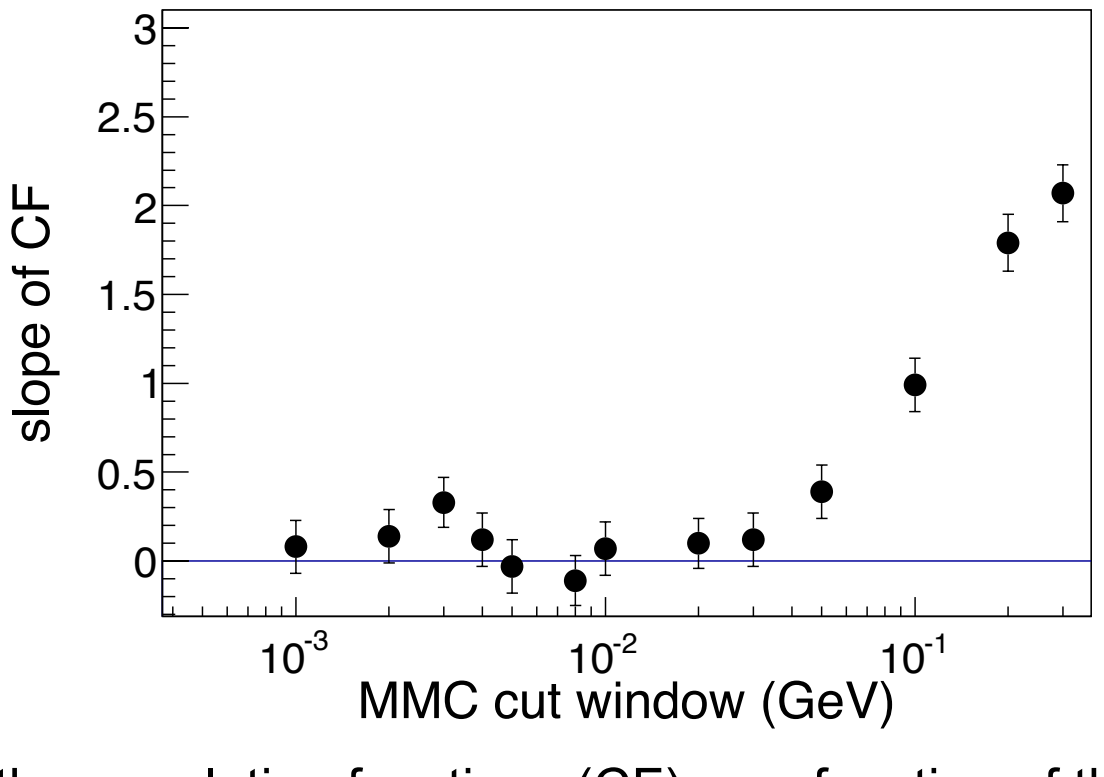

Fig. 4. The slope of the correlation functions (CF) as a function of the MMC cut window.

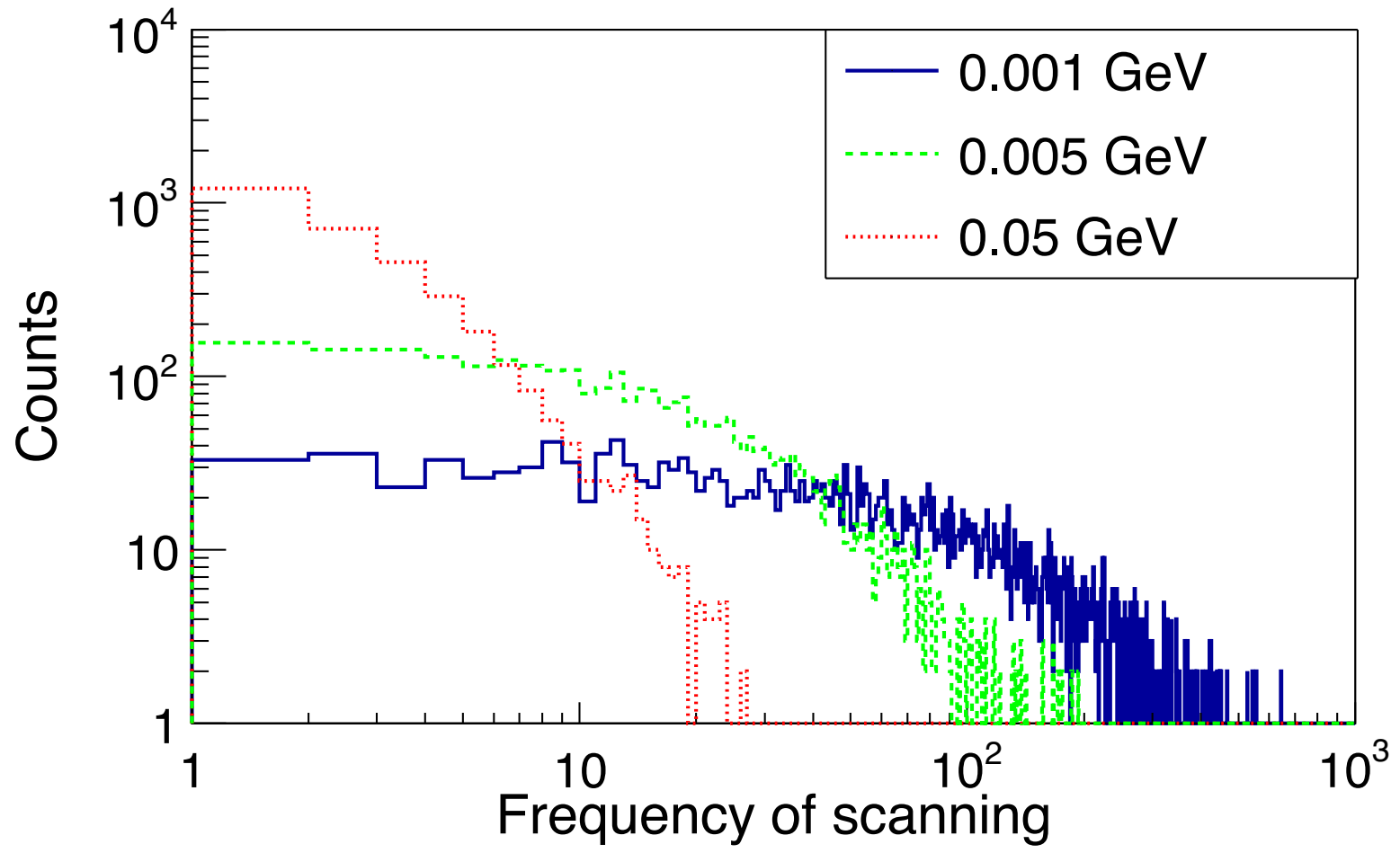

Fig. 5. Frequency of scanning in event mixing for the MMC windows of 0.001, 0.005, and 0.05 GeV.

We need to find an appropriate bin size of the incident photon energy to apply our event mixing method to real experimental data for the $\gamma p \rightarrow \pi^0\pi^0 p$ reaction. Although a wider bin size improves the statistics of mixed events, it would definitely influence the accuracy of the analysis. To test the effect of the energy bin size of the incident photons on the mixed events, we generate five samples of the $\gamma p \rightarrow \pi^0\pi^0 p$ event in the incident photon energy bins of 1.10-1.15 GeV, 1.11-1.15 GeV, 1.12-1.15 GeV, 1.13-1.15 GeV, and 1.14-1.15 GeV, respectively. The correlation functions obtained with event mixing for these five samples are shown in Fig. 6 (a). A linear fit is made for the correlation functions, indicating consistent results for these five different samples. A conclusion can be drawn that the energy bin size of 10 MeV is good enough for the analysis of experimental data obtained in the energy region between 1.10 and 1.15 GeV. When the energy bin width becomes wider, the slope of the correlation function increases slightly, as shown in Fig. 6 (b). Although the correlation function is almost flat even for the bin size 0.70-1.15 GeV, choosing wider bin size is not recommended in order to avoid unexpected problems.

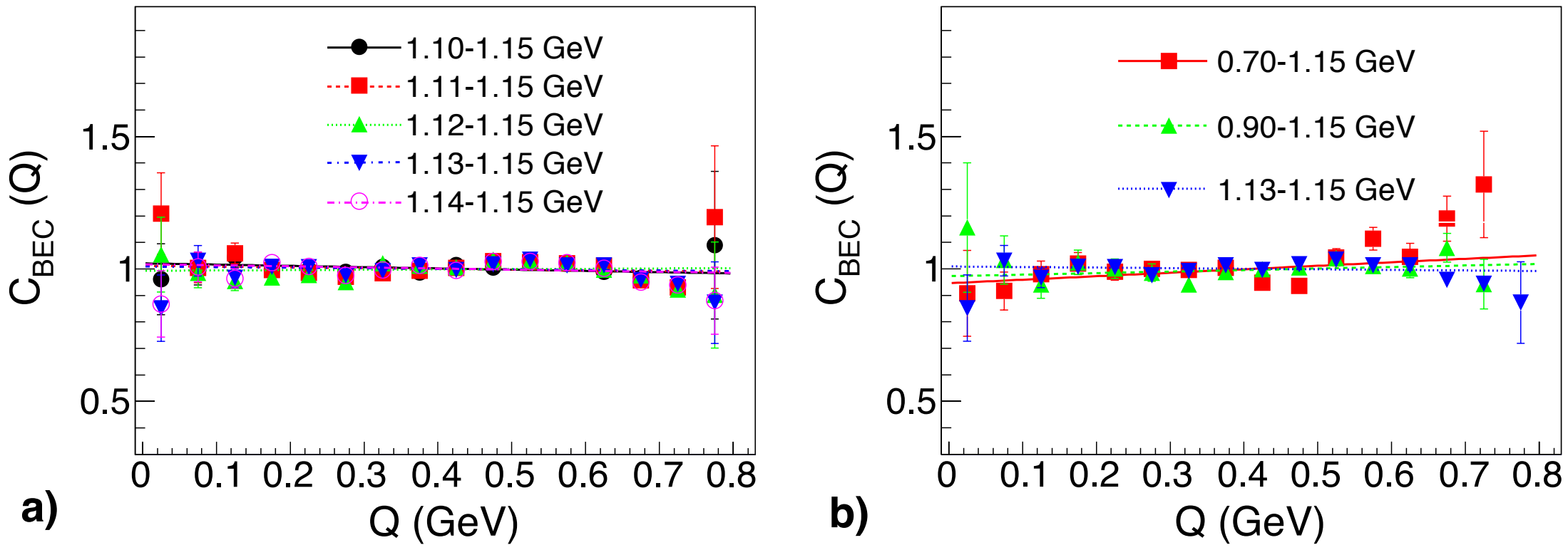

Fig. 6. (a) Correlation functions obtained with event mixing in multi-mixing mode (up to 10 times) for five samples of $\gamma p \rightarrow \pi^0\pi^0 p$ events in the incident photon energy bins of 1.10-1.15 GeV, 1.11-1.15 GeV, 1.12-1.15 GeV, 1.13-1.15 GeV, and 1.14-1.15 GeV, respectively. A linear function $y = aQ+b$ is fitted to the data (indicated by the straight lines). The five fit results are consistent with each other within error bars. (b) The same plot in the bins of 0.70-1.15 GeV, 0.90-1.15 GeV and 1.13-1.15 GeV.

To testify the ability of the multi mixing mode to observe the BEC effects, we use a BEC existent sample of the $\gamma p \rightarrow \pi^0\pi^0 p$ events with typical BEC parameters of $r_0$=0.8 fm and $\lambda_2$=0.5, which is produced by filtering a pure phase space event sample with a weight calculated by Eq.

(2) in terms of $Q$. Because Eq. (2) has a maximum value $C_{BEC}^{max} = N(1+\lambda_2)$ when $Q$ = 0, we employ the following procedure to construct a new sample with BEC effects from a pure phase space sample:

(i) Calculate the $Q$ value of two neutral pions in a phase-space event.

(ii) Generate a uniform random number $R$ ranging from 0.0 to 1.0.

(iii) Compare $R$ with the ratio $C_{BEC}(Q)/C_{BEC}^{max}$; if $C_{BEC}(Q)/C_{BEC}^{max} > R$ then accept this event.

The probability that $C_{BEC}(Q)/C_{BEC}^{max} > R$ is proportional to $C_{BEC}(Q)$, so this method can be used to produce a correct density distribution subject to Eq. (2).

In event mixing, both the single mixing mode and the multi-mixing mode (up to 10 mixed events for one original event) are employed. As shown in Fig. 7, these two modes yield consistent fit values of $r_0$ and $\lambda_2$ given in Eq. (2). These values are also in good agreement with the input BEC parameters of the generated sample. In practical applications to experimental data, it is important to investigate the BEC result dependence on the incident photon energy. Therefore, the same numerical tests for the $\gamma p \rightarrow \pi^0\pi^0 p$ events at six incident photon energies of $E_\gamma$=1.0 GeV, 1.03 GeV, 1.06 GeV, 1.09 GeV, 1.12 GeV, and 1.15 GeV are performed, showing the single and multiple mixing mode give consistent $r_0$ and $\lambda_2$ fit values, which are also consistent with the input BEC parameters of the generated samples (see Fig. 8). In addition, performing the same test at different photon energies provides more evidence that the multi mixing mode is feasible to observe BEC parameters. Good agreement between the multi and the single mixing modes results again proves the efficiency of the multi mode.

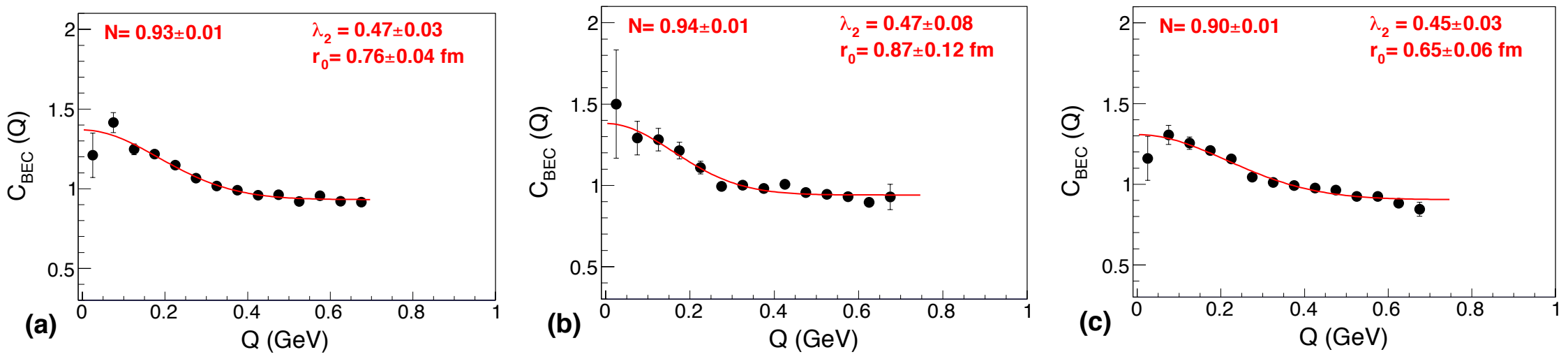

Fig. 7. Correlation functions for the generated sample (a), and mixed samples employing a single mixing mode (b) and a multi-mixing mode (c) for the $\gamma p \rightarrow \pi^0\pi^0 p$ events at an incident photon energy of 1.0 GeV. For the generated sample, the reference sample is from a pure phase space event sample free of BEC effects.

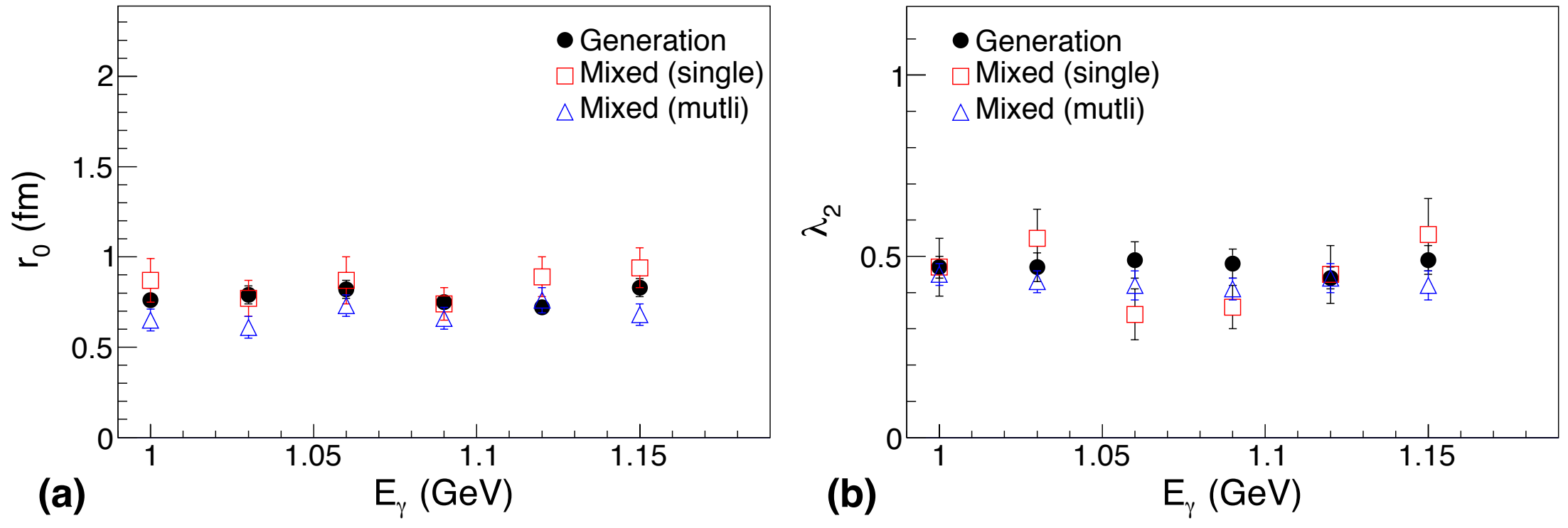

Fig. 8. Fitted values of $r_0$ (a) and $\lambda_2$ (b) obtained by event mixing in a single and a multi mixing (up to 10 times) mode at six incident photon energies $E_\gamma$=1.0, 1.03, 1.06, 1.09, 1.12, and 1.15 GeV for the $\gamma p \rightarrow \pi^0\pi^0 p$ events. For comparison, the values of $r_0$ and $\lambda_2$ for the generated sample with BEC effects are also shown.

## 4. Summary

Previous event mixing technique developed for inclusive reactions at high energies with a large multiplicity cannot be directly applied to exclusive reactions at low energies with a very limited multiplicity. Specific mixing constraints and mixing control conditions need to be developed in general to eliminate the factors induced by strict kinematical limits. An event mixing technique with a multi mixing mode is proposed for measuring two-pion Bose-Einstein correlations in the $\gamma p \rightarrow \pi^0\pi^0 p$ reaction around 1 GeV. The technique also employs a missing mass consistency (MMC) cut and a pion energy (PE) cut in order to effectively extract correlation parameters. The multi mixing allows an original event to be mixed with multi events and thus improves the statistics of the mixed event sample. Although the event mixing technique cannot reduce the genuine systematic uncertainty with the multi mixing mode, at least it can improve the statistical uncertainty for the reference sample. This feature improves the applicability of the event mixing technique. In numerical tests using $\gamma p \rightarrow \pi^0\pi^0 p$ events generated without BEC effects, the multi mixing mode is compared with the single mixing mode which requires one original event can be mixed only once. It is found the multi mixing mode can also make flat correlation function like the single mode. The multi mixing mode not only makes the results more reliable, but also provides an opportunity to cross check. The ability of the multi mixing mode to extract the BEC effects is also tested with samples having BEC effects. It is found the BEC parameters of $r_0$ and $\lambda_2$ obtained with multi event mixing are in good agreement with the input values within error bars.

**Acknowledgements**

This work was supported by the Institute of Fluid Physics, China Academy of Engineering, grant number 021600, and partially supported by the Ministry of Education and Science of Japan, Grant No. 19002003, and JSPS KAKENHI, Grant No. 24244022 and 26400287.